# Simulation and Optimization of MQW based optical modulator for on chip optical interconnect

**Sumita Mishra [1], Naresh k. Chaudhary[2] and Kalyan Singh[3]**

[1] E &C Engineering Department, Amity School of Engineering and Technology, Amity University
Lucknow, Uttar pradesh, India
*mishra.sumita@gmail.com*

[2] Electronics Department, Dr. RML Awadh University
Faizabad, Uttar pradesh, India
*Nkc3@rediffmail.com*

[3] Electronics Department, Dr. RML Awadh University
Faizabad, Uttar pradesh, India

**Abstract**

Optical interconnects are foreseen as a potential solution to improve the performance of data transmission in high speed integrated circuits since electrical interconnects operating at high bit rates have several limitations which creates a bottleneck at the interconnect level. The objective of the work is to model and then simulate the MQWM based optical interconnect transmitter. The power output of the simulated modulator is then optimized with respect to various parameters namely contrast ratio, insertion loss and bias current. The methodology presented here is suitable for investigation of both analog and digital modulation performance but it primarily deals with digital modulation. We have not included the effect of carrier charge density in multiple quantum well simulation.

***Keywords:*** *Optical interconnect, MQW, CR (contrast -ratio), IR (insertion loss).*

## 1. Introduction

The microprocessor industry has developed at an incredible pace, particularly in the past decades. Transistor scaling has been the crux of the rapid growth in processing power over the past forty years [1]. Scaling process has a large impact on electrical parameters of metallic interconnections which are responsible for transporting data within the microprocessor and between the microprocessor and memory and consequently, interconnect has become the dominant factor determining speed. Two fundamental interconnection limits encountered as the density of transistors increase one related to speed and the other to the number of input/output channels. Consequently as integrated circuit technology continues to scale if the interconnect problem is not addressed it will not be possible to achieve the exponential speed increases we have come to expect from the microprocessor industry. Optical interconnects have the potential to address this problem by providing both greater bandwidth and lower latency than electrical interconnects. Advantages offered by optical interconnects provide strong motivation to further develop methodologies for analysing optical interconnect links.

There have been several attempts at optimizing optical interconnect links using software tools such as Microsim P-Spice. [3-6]. P-Spice is designed for EDA and is not optimized for optical networks hence at very high frequencies precision of simulated circuit reduces. Thus for analysing the behaviour of high speed optical interconnects MATLAB and Simulink may be a more powerful tool since it offers multi-domain simulation environment and model-based design which can accurately model the behaviour of optical sub systems making it a good platform for optimization of optical interconnect link.

## 2. Background

For optical transmitters, VCSELs and MQWMs are the two primary optical sources for high density optical interconnects. However VCSEL's use is limited due to self-heating and device lifetime concerns [5] .

Quantum-well modulators have so far been the devices most extensively used in demonstrating actual dense interconnects to and from silicon CMOS chips. [6,9] These devices have successfully been made in large arrays and solder bonded to the circuits. Also, Multiple quantum well (MQW) modulators offer an advantage over other light emitters in terms of signal and clock distribution. Furthermore, the electrical signals can be sampled with short optical pulses to improve the performance of receivers. MQWM based link requires that an external beam be brought onto the modulator. This facilitates to generate and control one master laser beam which allows centralized clocking of the entire system, and the use of modulators, as described above, allows the retiming of signals, especially if the master laser operates with relatively short optical pulses. Thus QWM based approach, besides yielding lower transmitter on-chip power dissipation can be more conducive to monolithic integration. This was the motivation for simulating a MQWM based optical interconnect link.

## 3. Modeling and Simulation methodology

In this section we describe the methodology used for modelling and simulation of optical interconnect transmitter.

The simulated laser diode is an InGaAs–Al-GaAs–GaAs quantum-well separate confinement heterostructure. We considered only the internal parasitics assuming a low-parasitics assembly scheme. The simulated modulator structure is reflective mode (RMQWM).

For simulation of the dynamic response of MQW laser a rate equation model has been used [7]. In this model we have not included the effect of carrier dynamics in the quantum wells yielding the following set of equations

$$\frac{dN(t)}{dt} = \frac{I(t)}{qV_a} - g_0 \frac{N(t) - N_0}{1 + \varepsilon S(t)} S(t) - \frac{N(t)}{\tau_n} \quad (1)$$

$$\frac{dS(t)}{dt} = g_0 \Gamma \frac{N(t) - N_0}{1 + \varepsilon S(t)} S(t) - \frac{S(t)}{\tau_p} + \Gamma \beta \frac{N(t)}{\tau_n} \quad (2)$$

$$\frac{d\phi}{dt} = \frac{\alpha}{2}\left[\Gamma g_0 [N(t) - N_0] - \frac{1}{\tau_p}\right] \quad (3)$$

With

$$p(t) = \frac{S(t)\eta h\nu V_a}{2\Gamma\tau_p} \quad (4)$$

Where N(t) is a the carrier density in the in the quantum wells, S is the photon density in the laser cavity, $\phi$ is the phase of the optical field, I is the injection current, q is the electronic charge, $N_0$ is the carrier density in the quantum wells for the reference bias level, p is the power output. physical meaning and values of various other coefficients can be found in ref [7]. Simulated Laser power output was then fed to the modelled integrated surface-normal reflective electroabsorption mqw modulators. Quantum well absorption data for three quantum wells is taken from the literature for well width of 95 Å, and the Al0.3Ga0.7As barrier thickness of 30 Å. An electroabsorption modulator using the quantum-confined Stark effect is formed by placing an absorbing quantum well region in the intrinsic layer of a pin diode. Doing so creates the typical p-i-n photodiode structure and enables large fields to be placed across the quantum wells without inducing large currents. By applying a static reverse bias across the diode, photogenerated carriers are efficiently swept out of the intrinsic region and the device acts as a photodetector. Varying this bias causes a modulation in the optical absorption, resulting in an optical modulator. The modulator is characterized by its capacitance, Insertion Loss and Contrast Ratio. An ideal modulator has minimum optical power loss during the "on" state (IL), and largest possible optical power ratio between the "on" and the "off" states (CR). Typically, there is a trade-off between these parameters for a given value of the ratio between maximum (αmax) and minimum (αmin) absorption. The IL/CR relation for a simple RMQW structure in a reverse biased PIN configuration is given below

$$CR = \frac{R_{on}}{R_{off}} = \frac{e^{-2\alpha_{min} l}}{e^{-2\alpha_{max} l}} = (1 - IL)^{1-x} \quad (5)$$

Here $R_{on}$ and $R_{off}$ are the modulator reflectivities in the less absorbing and more absorbing states, respectively. CR decreases significantly at low operating voltages.

The modulator power output consists of the dynamic component including the capacitance of the driver chain and the modulator and the static component due to the absorbed optical power in the "on" and the "off" state. At low voltages, the dynamic component is small. The static power is calculated in terms of the CR and IL by multiplying the current in each binary state by its respective voltages and taking an average [9]

$$P = .5 R_{s\,mod} P_l [IL(V_{bias} - V_{dd}) + (1 - \frac{1-IL}{CR})V_{bias}] = \eta_{mod} P_l \quad (6)$$

Here $\eta_{mod}$ is a dimensionless efficiency factor $R_{s\,mod}$ is the modulator responsively $P_l$ is the input laser power to the modulator, $V_{bias}$ the pre-bias voltage and $V_{dd}$ is the supply voltage small compared to the static power of the modulator.

## 4. Model description and Results

Simulation was carried out in two stages. In the first stage the rate equations were implemented in simulink as shown in fig -1. Laser power output was then coupled to external modulator. Simulink model of MQWM modulator is shown in fig 2. Simulated Laser diode photon density for 1ns pulse is shown in Figure 3. The simulated power output response of MQWM modulator is shown in fig 4. Simulated optical photon density output of MQWM Modulator with ramp input and bias current=2mA is shown in fig 5. Minimun interconnect power is observed as a function of bit rate. We further study the change in the minimum interconnect power as a function of parameter X, which is dictated by bias current. It was observed that response of model worsens with increase in bias current. We have not included the effect of pattern jitters and crosstalk. All the simulations were run over a time period that was several orders of magnitude longer than the fixed step size chosen so that turn-on transient effects that happen near threshold can be avoided. All simulations were carried out using standard 4th-order Runge-Kutta algorithm with a fixed step size.

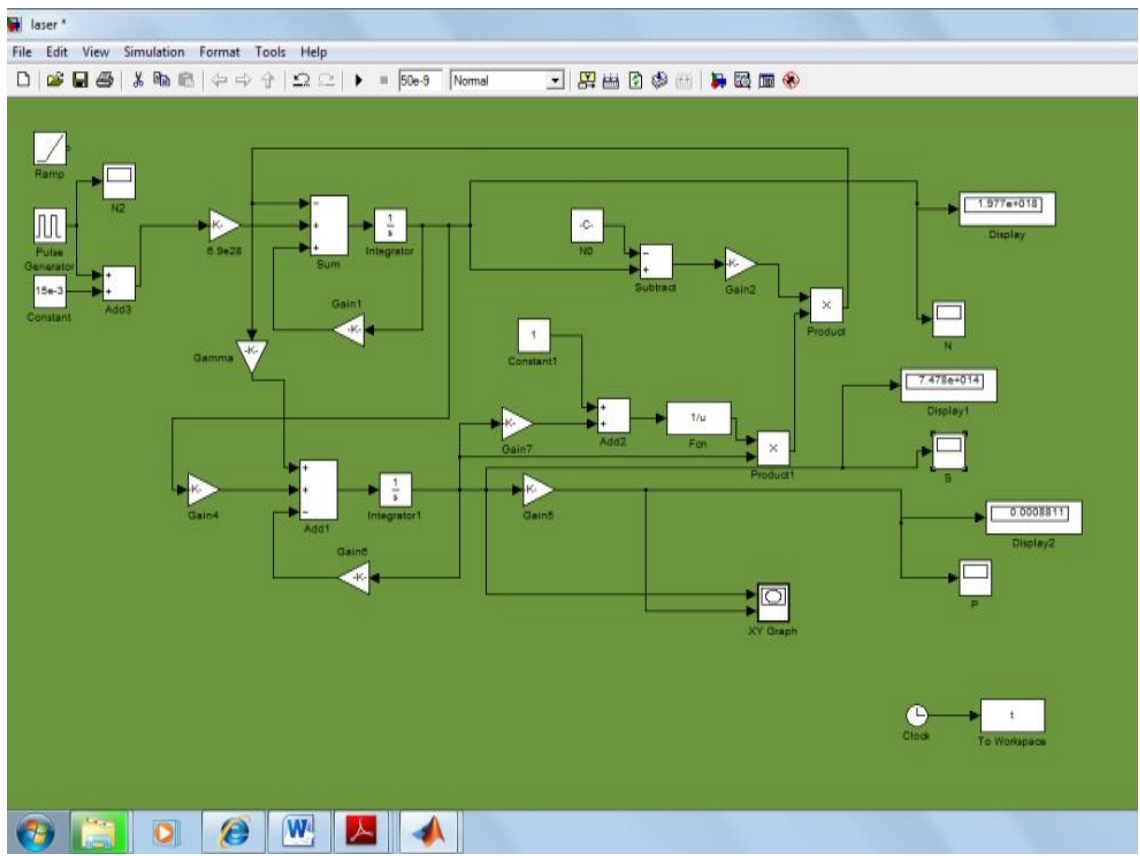

**Figure 1. Simulink model of LASER**

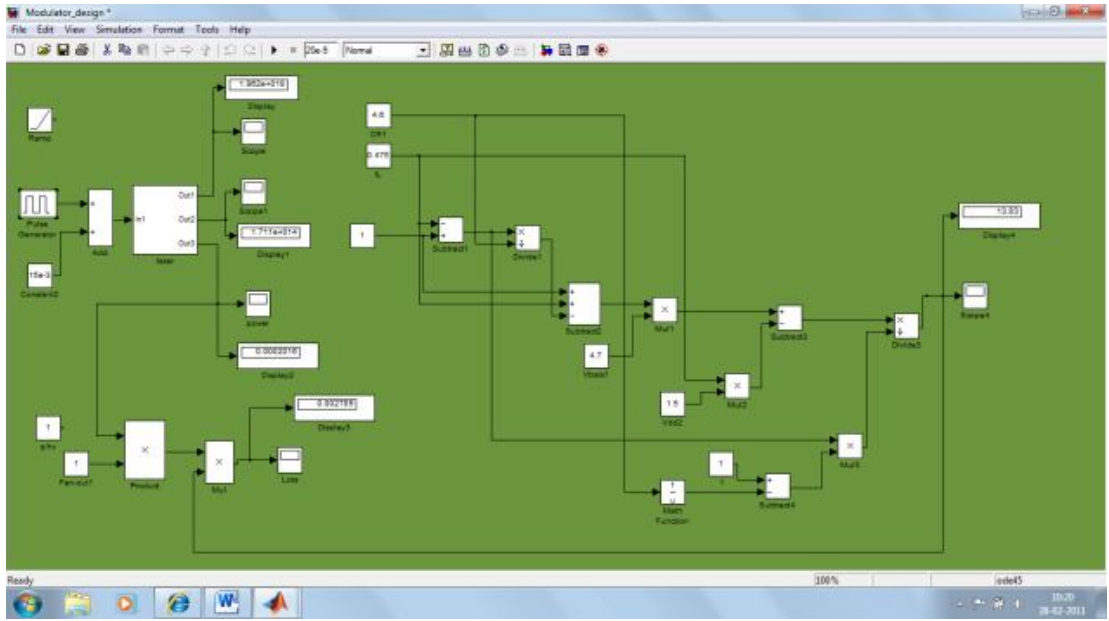

**Figure 2. Simulink model of Optical Modulator**

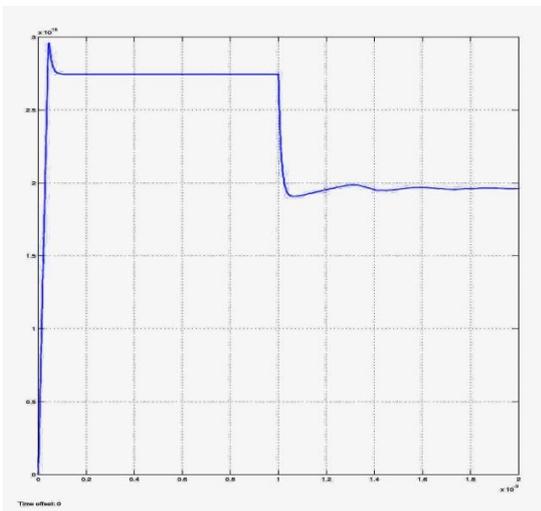

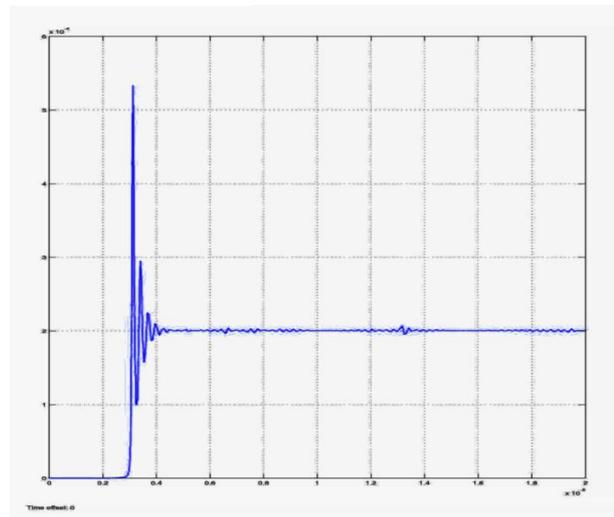

**Fig.4 Power Dissipation of MQWM**

**Fig.3 Simulated photon density of Laser with bias current=1mA**

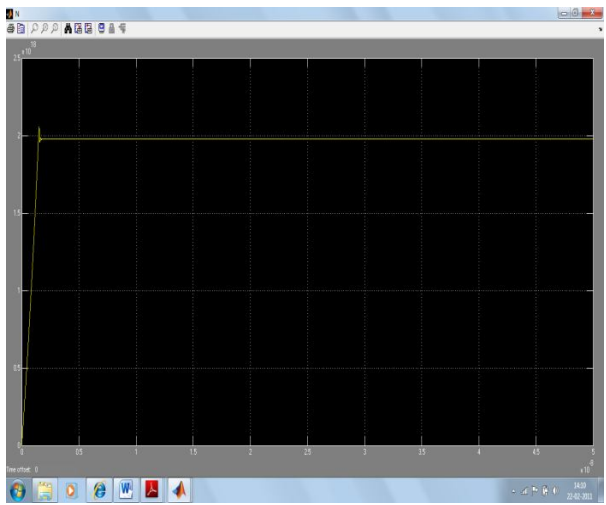

Fig. 5 Simulated photon density of MQWM Modulator with ramp input and bias current=2mA

## 5. Conclusions

The work describes a methodology to model, simulate and then optimize the MQWM based optical interconnect transmitter power output with respect to various parameters namely contrast ratio, insertion loss and bias current. The methodology presented here is suitable for investigation of both analog and digital modulation performance but it primarily deals with digital modulation. The modulator was simulated on MATLAB Simulink tool and model response was obtained for 1- 20Gbps bit rate. The simulated model can achieve error-free operation under 16 Gbps data rate. It was observed that Modulator output worsens with increase in bias current. These results are based on simplified cases excluding pattern jitters, crosstalk and the effect of carrier charge density in multiple quantum well. However, the effect of pattern jitters and bandwidth limits of each device will become increasingly important as the density of an interconnect array becomes higher. These are subjects for further study. The model can be further improved by addressing these issues.

**Acknowledgments**

The first author Sumita Mishra is grateful to Maj. Gen. K.K. Ohri , Prof. S.T.H. Abidi and Brig. U. K. Chopra of Amity University, Lucknow  for their support during the research work.

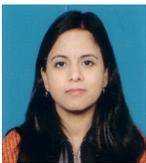

**Sumita Mishra** has done her post-graduation in Electronics Science from Lucknow University, Lucknow. She has also done M. Tech (Optical Communication) from SGSITS, Indore in 2004. Thereafter, she was appointed as a lecturer in Electronics and Communication Engineering Department at Amity School of Engineering and Technology (New Delhi). Currently, She is working as a lecturer in ECE department at Amity University (Lucknow campus), she is also pursuing doctoral degree in Electronics at DRML Awadh University, India. She is a member of IEC, Oxford Journals, ABI Research, Transmission & Distribution World IEEE, ACM, IEEE (institutional membership), PCPro, IACSIT and VLSI Jagrati. Her current research interests include Fibre Optic CDMA, Optical interconnects and Machine vision. Her research papers (6) have been presented in various IEEE international and National conferences. She has two publications in international journals.

**Naresh K Chaudhary** has done his Ph.D in Physics from Lucknow University, Lucknow in 2005, Thereafter, he joined as a lecturer in Institute of Engineering and Technology Resura Sitapur UP INDIA. Where he worked till November 2006 .Then he was appointed as Assistant professor in Department of Electronics and Physics at Dr RML Avadh University Faizabad in Nov 2006. He has seven publications in various national and international journals.

**Kalyan singh** Dr Kalyan Singh is Professor and head Department of Physics and Electronics Dr RML Avadh University Faizabad. He has 37 year of post Ph.D experience .He has guided 11 Research Scholars and published 26 papers at national and international level.